\documentclass[english]{article}

\usepackage[T1]{fontenc}
\usepackage[latin9]{inputenc}
\usepackage{amsthm,amsmath,amssymb}
\usepackage{babel}
\usepackage{fullpage}

\begin{document}

\newtheorem{thm}{Theorem}[section]
\newtheorem{cor}{Corollary}[section]
\newtheorem{lem}{Lemma}[section]
\newtheorem{prop}{Proposition}[section]

\def\const{\text{const.}}
\def\Rnum{{\mathbb R}}
\def\sgn{{\rm sgn}}
\def\Dop{{\mathcal{D}}}
\def\nvec{{\mathbf n}}
\def\kvec{{\mathbf k}}
\def\sech{{\rm sech}}
\def\sig{\sigma^2}
\def\sqrtsig{\sigma}

\allowdisplaybreaks[2]

\title{Conservation laws, symmetries, and line soliton\\ solutions of generalized KP and Boussinesq\\ equations with $p$-power nonlinearities\\ in two dimensions}

\author{S.C. Anco${}^1$, M.L. Gandarias${}^2$, E. Recio${}^2$
\\${}^1$Brock University, St Catharines Canada 
\\${}^2$Cadiz University, Cadiz  Spain}

\date{}
\maketitle

\begin{abstract}
Nonlinear generalizations of integrable equations in one dimension, 
such as the KdV and Boussinesq equations with $p$-power nonlinearities, 
arise in many physical applications and are interesting in analysis due to critical behaviour. 
This paper studies analogous nonlinear $p$-power generalizations of the integrable KP equation and the Boussinesq equation in two dimensions. 
Several results are obtained. 
First, for all $p\neq 0$, a Hamiltonian formulation of both generalized equations is given. 
Second, all Lie symmetries are derived, 
including any that exist for special powers $p\neq0$. 
Third, Noether's theorem is applied to obtain the conservation laws arising from 
the Lie symmetries that are variational. 
Finally, explicit line soliton solutions are derived for all powers $p>0$,
and some of their properties are discussed.  
\end{abstract}

\section{Introduction}

The Kadomtsev-Petviashvili (KP) equation 
\begin{equation}\label{KP}
(u_t+\alpha u u_x +u_{xxx})_x+ \beta u_{yy}=0
\end{equation}
and its modified version 
\begin{equation}\label{mKP}
(u_t+\alpha u^2 u_x \pm\sqrt{2\alpha|\beta|}u_x\partial_x^{-1}u_y+u_{xxx})_x+\beta u_{yy}=0
\end{equation}
are integrable systems in $2+1$ dimensions. 
These equations arise in important physical applications, 
most notably as models for shallow water waves \cite{AblSeg}, ion acoustic waves \cite{KadPet}, and magnetic excitations in thin films \cite{VeeDan}, 
and they possess line soliton and lump solutions \cite{Sat,SatAbl,ManZakBorMat,GesHolSaaSim,KonDub}.
Both equations have a rich mathematical structure. 
In particular, their dimensional reductions yield many physically relevant integrable systems in $1+1$ dimensions. 

Nonlinear generalizations of $1+1$ dimensional integrable systems 
with $p$-power nonlinearities have been extensively studied. 
The best known example is the generalized Korteweg-de Vries (KdV) equation 
$u_t + u^p u+ u_{xxx}=0$, with $p\neq0$, 
which becomes the ordinary KdV equation when $p=1$ 
and the modified KdV equation when $p=2$. 
The KdV equation is a model of uni-directional shallow water waves,
and the modified KdV equation models uni-directional ion acoustic waves. 
For a general power $p$, 
the generalized KdV equation has solitary wave solutions 
$u = (\tfrac{1}{2}(p+1)(p+2)c)^{1/p} {\rm sech}^{2/p}\big( \tfrac{p}{2}\sqrt{c} (x-ct) \big)$
whose interactions depend sensitively on the value of $p$. 
This sensitivity can be understood by looking at the scaling properties of 
the conserved mass $\mathcal{M}[u]= \int_{-\infty}^{\infty} u^2\,dx$ 
and the conserved energy 
$\mathcal{E}[u]= \int_{-\infty}^{\infty}\big( \tfrac{1}{2}u_x^2 -\tfrac{1}{(p+1)(p+2)}u^{p+2} \big)\,dx$ for initial-value solutions $u(x,t)$. 
Under the scaling symmetry $x\to\lambda x$, $t\to \lambda^3 t$, $u\to\lambda^{-2/p}u$, 
the mass is scaling-invariant when $p=4$, 
while the energy has a negative scaling weight for any $p>0$. 
These properties can be used to show \cite{Tao} that solitary waves are stable 
under arbitrary perturbations with respect to the $H^1$ norm, 
$||u||_{H^1} = \int_{-\infty}^{\infty}\big( u^2 + u_x^2 \big)\,dx$, 
provided that $p$ is smaller than the critical power $p^*=4$. 
This stability extends to the critical case $p=4$ for solitary waves of sufficiently small energy. 

There has been much less study of $p$-power generalizations of the KP and modified KP 
equations themselves. 
In particular, 
the generalized KP (gKP) equation
\begin{equation}\label{gKP}
(u_t+\alpha u^p u_x +u_{xxx})_x+ \beta u_{yy}=0, 
\quad p\neq 0 
\end{equation}
is a natural $2+1$ dimensional analog of the gKdV equation. 
An interesting variant is given by 
\begin{equation}\label{gB}
u_{tt}=u_{xx}+\alpha (u^{p+1})_{xx}+\gamma u_{xxxx}+\beta u_{yy},
\quad p\neq 0 
\end{equation} 
which is a $2+1$ dimensional generalized Boussinesq (2D gB) equation. 
Its reduction to $1+1$ dimensions gives a $p$-power generalization of 
the ordinary Boussinesq equation 
\begin{equation}\label{Bousi}
u_{tt}=u_{xx}+\alpha (u^2)_{xx}+ \gamma u_{xxxx} . 
\end{equation} 
This equation is an integrable system 
which can be obtained by dimensional reduction of the KP equation, 
and it is a physically relevant model of bi-directional shallow water waves. 
One very interesting feature of the Boussinesq equation when $\gamma<0$ is that 
its solitary wave solutions have an instability 
which can lead to formation of a singularity in a finite time \cite{BogZak,FalSpeTur}. 
This behaviour is in contrast to the typical stability property of solitons in integrable systems. 

The purpose of the present paper is to determine 
the conservation laws, symmetries, and line soliton solutions 
admitted by the gKP equation \eqref{gKP} and the 2D gB equation \eqref{gB} 
in $2+1$ dimensions, for all nonlinearity powers $p>0$. 

First, in section~\ref{hamiltonian}, 
both the gKP and 2D gB equations for arbitrary $p\neq0$ are formulated as Hamiltonian systems
by the introduction of potentials. 
Next, all Lie point symmetries are obtained in section~\ref{symms},
and the class of variational point symmetries is found. 
Through Noether's theorem in the Hamiltonian setting, 
the corresponding conservation laws are derived in section~\ref{conslaws}. 
These results include any point symmetries and any Noether conservation laws 
that arise for special powers $p\neq0$. 

In section~\ref{solns}, 
line solitons $u=U(x+\mu y-\nu t)$ are considered for both the gKP and 2D gB equations for all $p>0$, 
where the parameters $\mu$ and $\nu$ determine the direction and the speed of the line soliton. 
The resulting fourth-order nonlinear differential equations for $U$
are directly reduced to first-order separable differential equations
by use of the conservation laws obtained for the gKP and 2D gB equations. 
These separable differential equations are then integrated to get the explicit form of 
the line soliton solutions for all $p>0$, 
with $\mu$ and $c$ being arbitrary apart from satisfying a kinematic inequality. 
Some properties of these solutions are discussed. 

Finally, a few concluding remarks are made in section~\ref{remarks}. 

A few conservation laws of the gKP equation \eqref{gKP} in the case $p=2$
and the 2D gB equation \eqref{gB} in the case $p=2$ 
have been found previously in Refs.\cite{NazAliNae,RuiZhaZha}. 
Some solutions of the gKP equation in three dimensions, without any discussion of their properties, are obtained in Ref.\cite{AdeKhaBis}
A general treatment of symmetries, conservation laws, 
and their applications to differential equations
can be found in Refs.\cite{Olv,BluCheAnc}.

\section{Hamiltonian structure}\label{hamiltonian}

It is useful to begin by noting that 
the parameters in the gKP equation \eqref{gKP} transform 
under scalings 
$t\to\lambda^3 t$, $x\to \lambda x$, $y\to\lambda^a y$, $u\to \lambda^b u$  
as $\alpha\to \lambda^{-pb-2}\alpha$ and $\beta\to \lambda^{2a-4}\beta$. 
Hence, without loss of generality, 
we can put $\alpha =1$ and $\beta=\pm1$ in the gKP equation \eqref{gKP}. 
We can likewise put $\alpha =1$, $\beta=\pm1$, $\gamma=\pm1$ in the 2D gB equation \eqref{gB}. 
Thus, we will hereafter consider the gKP and gB equations in the respective forms
\begin{equation}\label{gkp}
(u_t+u^p u_x +u_{xxx})_x +\sig u_{yy}=0,
\quad p\neq 0 
\end{equation}
and 
\begin{equation}\label{gb}
u_{tt}=u_{xx}+ (u^{p+1})_{xx} \pm u_{xxxx}+\sig u_{yy},
\quad p\neq 0 
\end{equation} 
where $\sig=\pm 1$. 

Both the gKP equation \eqref{gkp} and the 2D gB equation \eqref{gb}
can be expressed as Hamiltonian evolution equations. 
The simplest way to derive this formulation is by first considering 
a Lagrangian form of the equations. 

For the gKP equation, 
a Lagrangian is known for the case $p=1$ \cite{Dic}
by the use of a potential $v(x,y,t)$ given by 
\begin{equation}\label{pot}
u=v_x .
\end{equation}
In terms of this potential, 
the gKP equation has the form 
\begin{equation}\label{gkp-pot}
0 = v_{tx} +v_x^p v_{xx} +v_{xxxx} +\sig v_{yy}
= \delta L/\delta v
\end{equation}
where $L$ is the Lagrangian 
\begin{equation}\label{gkp-L}
L = -\tfrac{1}{2}v_tv_x +\tfrac{1}{2}v_{xx}^2 - \tfrac{1}{(p+1)(p+2)}v_x^{p+2} -\tfrac{1}{2}\sig v_y^2 .
\end{equation}
The associated Hamiltonian formulation is then given by 
\begin{equation}\label{gkp-pot-hamil}
v_t = -D_x^{-1}(\delta H/\delta v)
\end{equation}
with 
\begin{equation}\label{gkp-H}
H = \int_{\Rnum^2} \big( \tfrac{1}{2}v_{xx}^2 -\tfrac{1}{(p+1)(p+2)}v_x^{p+2} -\tfrac{1}{2}\sig v_y^2  \big)\,dx\,dy
\end{equation}
being the Hamiltonian functional,
and with $D_x^{-1}$ being a Hamiltonian operator \cite{Olv}. 
This formulation can be equivalently expressed in terms of $u$ 
by using the variational derivative identity 
$\delta H/\delta v = -D_x(\delta H/\delta u)$,
which yields
\begin{equation}\label{gkp-hamil}
u_t = D_x(\delta H/\delta u)
\end{equation}
where the Hamiltonian density is now a nonlocal expression in terms of $u$. 
Note that if we put $\sig=0$, 
then we obtain the standard Hamiltonian formulation of the gKdV equation \cite{AncRecGanBru}.

A similar Lagrangian formulation can be obtained for the 2D gB equation 
starting from the same potential \eqref{pot}. 
The corresponding form of the 2D gB equation is given by 
\begin{equation}\label{gb-pot}
0= v_{tt}-v_{xx} -(v_x^{p+1})_x \mp v_{xxxx} -\sig v_{yy}
= \delta L/\delta v
\end{equation}
with 
\begin{equation}\label{gb-L}
L = -\tfrac{1}{2} v_t^2 +\tfrac{1}{2}v_x^2 + \tfrac{1}{p+2}v_x^{p+2} \mp \tfrac{1}{2}v_{xx}^2 +\tfrac{1}{2}\sig v_y^2 
\end{equation}
being the Lagrangian. 
To obtain the associated Hamiltonian formulation, 
it is necessary to convert this formulation into an equivalent first-order evolution system
\begin{equation}\label{gb-potsys}
v_t = w,
\quad
w_t = v_{xx} + (v_x^{p+1})_x \pm v_{xxxx} +\sig v_{yy} . 
\end{equation}
The associated Hamiltonian formulation for this system is then given by 
\begin{equation}\label{gb-potsys-hamil}
\begin{pmatrix}
v\\ w
\end{pmatrix}_t 
= 
J \begin{pmatrix}
\delta H/\delta v \\\delta H/\delta w
\end{pmatrix},
\quad
J= \begin{pmatrix}
0 & 1\\ -1 & 0
\end{pmatrix}
\end{equation}
where 
\begin{equation}\label{gb-H}
H = \int_{\Rnum^2} \big( 
\tfrac{1}{2}w^2 +\tfrac{1}{2}v_x^2 + \tfrac{1}{p+2}v_x^{p+2} \mp \tfrac{1}{2}v_{xx}^2 +\tfrac{1}{2}\sig v_y^2 \big)\,dx\,dy
\end{equation}
is the Hamiltonian functional,
and where $J$ plays the role of a Hamiltonian operator \cite{Olv}. 
An equivalent formulation in terms of $u$ arises from the relation
$u_t = v_{tx} = w_x$ 
combined with the previous variational derivative identity $\delta H/\delta v = -D_x(\delta H/\delta u)$. 
This yields 
\begin{equation}\label{gb-hamil}
\begin{pmatrix}
u\\ w
\end{pmatrix}_t 
= 
\Dop \begin{pmatrix}
\delta H/\delta u \\\delta H/\delta w
\end{pmatrix},
\quad
\Dop= \begin{pmatrix}
0 & D_x\\ D_x & 0
\end{pmatrix}
\end{equation}
with $\Dop$ being a Hamiltonian operator \cite{Olv},
and with the Hamiltonian density now being a nonlocal expression in terms of $u,w$. 
Note that if we put $\sig=0$, 
then we obtain a Hamiltonian formulation of the 1D gB equation, 
which coincides in the case $p=1$ 
with the first Hamiltonian structure \cite{Olv} of the ordinary Boussinesq equation \eqref{Bousi}. 

These Hamiltonian formulations motivate studying 
the symmetries and conservation laws of the gKP and 2D gB equations 
in their respective potential forms \eqref{gkp-pot} and \eqref{gb-pot}.

\section{Symmetries}\label{symms}

Symmetries are a basic structure of nonlinear evolution equations. 
They yield transformation groups under which the set of solutions of a given evolution equation is mapped into itself, 
and thus they can be used to find group-invariant solutions. 

We will first consider point symmetries $(x,y,t,v) \to (\tilde x,\tilde y,\tilde t,\tilde v)$,
where $\tilde x,\tilde y,\tilde t,\tilde v$ are functions of $x,y,t,v$. 
For the gKP and 2D gB equations \eqref{gkp-pot} and \eqref{gb-pot}, 
an infinitesimal point symmetry consists of a generator
\begin{equation}
\mathrm{X} =\xi^x(x,y,t,v)\partial_x+\xi^y(x,y,t,v)\partial_y+\tau(x,y,t,v)\partial_t+\eta(x,y,t,v)\partial_v
\end{equation}
whose prolongation leaves invariant the respective equation. 
Every point symmetry can be expressed in an equivalent characteristic form
\begin{equation}
\mathrm{\hat X} =P\partial_v,
\quad
P=\eta(x,y,t,v)-\xi^x(x,y,t,v)v_x-\xi^y(x,y,t,v)v_y-\tau(x,y,t,v)v_t,
\end{equation}
which acts only on $v$.
The function $P$ is called the symmetry characteristic. 

Invariance of the gKP equation in potential form \eqref{gkp-pot} 
is expressed by the condition
\begin{equation}
0=D_x(D_t P + v_x^p D_x P +D_3^4 P)  +\sig D_y^2 P
\end{equation}
holding for all solutions $v(x,y,t)$ of the gKP potential equation.
Similarly, invariance of the 2D gB equation in potential form \eqref{gb-pot} is given by 
\begin{equation}
0=D_t^2 P -D_x^2( P +(p+1) v_x^p D_xP \pm D_x^2 P)  -\sig D_y^2 P
\end{equation}
holding for all solutions $v(x,y,t)$ of the 2D gB potential equation.
Each of these invariance conditions 
splits with respect to $x$-derivatives and $y$-derivatives of $v$,
yielding an overdetermined system of equations on 
$\eta(x,y,t,v)$, $\xi^x(x,y,t,v)$, $\xi^y(x,y,t,v)$ $\tau(x,y,t,v)$, along with  $p$,
subject to the classification condition $p\neq 0$.
It is straightforward to set up and solve these two determining systems by using Maple. 
In particular, 
the Maple package 'rifsimp' can be used to obtain a complete classification of all solution cases. 

We will now summarize the results.

\begin{thm}\label{pointsymms-gkp}
(i) 
The point symmetries admitted by the generalized KP potential equation \eqref{gkp-pot}
for arbitrary $p\neq 0$ 
are generated by
\begin{align}
& 
\tau_1 = 0,
&&
\xi_1^x = 1,
&&
\xi_1^y = 0,
&&
\eta_1 = 0 ;
\label{gkp-symm1}
\\
&
\tau_2 = 0,
&&
\xi_2^x = 0,
&&
\xi_2^y = 1,
&&
\eta_2 = 0 ;
\label{gkp-symm2}
\\
&
\tau_3 = 1,
&&
\xi_3^x = 0,
&&
\xi_3^y = 0,
&&
\eta_3 = 0 ;
\label{gkp-symm3}
\\
&
\tau_4 = 3t,
&&
\xi_4^x = x,
&&
\xi_4^y = 2y,
&&
\eta_4 = (1-\tfrac{2}{p})v ;
\label{gkp-symm4}
\\
&
\tau_5 = 0,
&&
\xi_5^x = y,
&&
\xi_5^y = -2\sig t,
&&
\eta_5 = 0 ;
\label{gkp-symm5}
\\
&
\tau_6 = 0,
&&
\xi_6^x = 0,
&&
\xi_6^y = 0, 
&&
\eta_6 = f_1(t)+f_2(t)y .
\label{gkp-symm6}
\end{align}
Symmetries \eqref{gkp-symm1}--\eqref{gkp-symm3} are translations, 
symmetry \eqref{gkp-symm4} is a scaling, 
and symmetry \eqref{gkp-symm5} is a rotation in the plane $(y,t+\sig x)$,
combined with a boost in the plane $(y,t -\sig x)$. 
Symmetry \eqref{gkp-symm6} is a linear combination of $2$ infinite-dimensional families,
corresponding to the general solution of $P_{yy}=0$ for $P(y,t)$.
(ii) 
Additional point symmetries are admitted by the generalized KP potential equation \eqref{gkp-pot}
only when $p=1$:
\begin{equation}\label{gkp-symm7}
\begin{aligned}
&
\tau_7 = f_3(t),
\quad
\xi_7^x = -\tfrac{1}{6}\sig f''_3(t)y^2 -\tfrac{1}{2}\sig f'_4(t)y+\tfrac{1}{3}f'_3(t)x+f_5(t),
\\
&
\xi_7^y = \tfrac{2}{3}f'_3(t)y +f_4(t),
\\
&
\eta_7 = \tfrac{1}{72}f''''_3(t)y^4  + \tfrac{1}{12}\sig f'''_4(t)y^3 
-\tfrac{1}{6}\sig f'''_3(t)y^2x -\tfrac{1}{2}\sig f''_5(t)y^2 
-\tfrac{1}{2}f''_4(t)yx 
\\&\qquad
+\tfrac{1}{6}f''_3(t)x^2+f'_5(t)x -\tfrac{1}{3}f'_3(t)v.
\end{aligned}
\end{equation}
This symmetry is a linear combination of $3$ infinite-dimensional families
and includes the $p=1$ case of the scaling symmetry \eqref{gkp-symm4} 
for $f_3(t)=3t$, $f_4(t)=f_5(t)=0$. 
\end{thm}

The corresponding symmetry transformation groups for arbitrary $p\neq0$ are respectively given by 
\begin{align}
(x,y,t,v)\to & 
(x+\epsilon,y,t,v) , 
\\
(x,y,t,v)\to &
(x,y+\epsilon,t,v) , 
\\
(x,y,t,v)\to &
(x,y,t+\epsilon,v) , 
\\
(x,y,t,v)\to &
(e^{\epsilon}x,e^{2\epsilon}y,e^{3\epsilon}t,e^{(1-2/p)\epsilon}v) , 
\label{gkp-scaling}
\\
(x,y,t,v)\to &
(x+\epsilon y-\sig\epsilon^2 t,y-2\epsilon\sig t,t,v) , 
\\
(x,y,t,v)\to &
(x,y,t,f_1(t)\epsilon+f_2(t)\epsilon y+v) ,
\end{align}
where $\epsilon\in\Rnum$ is the group parameter. 
For $p=1$, 
the symmetry transformation group generated by the three infinite families \eqref{gkp-symm7}
is discussed in Ref.\cite{Lou}. 

\begin{thm}\label{pointsymms-gb}
(i) 
The point symmetries admitted by the 2D generalized Boussinesq potential equation \eqref{gb-pot}
for arbitrary $p\neq 0$ 
are generated by
\begin{align}
& 
\tau_1 = 0,
&&
\xi_1^x = 1,
&&
\xi_1^y = 0,
&&
\eta_1 = 0 ;
\label{gb-symm1}
\\
& 
\tau_2 = 0,
&&
\xi_2^x = 0,
&&
\xi_2^y = 1,
&&
\eta_2 = 0 ;
\label{gb-symm2}
\\
& 
\tau_3 = 1,
&&
\xi_3^x = 0,
&&
\xi_3^y = 0,
&&
\eta = 0 ;
\label{gb-symm3}
\\
&
\tau_4 = y,
&&
\xi_4^x = 0,
&&
\xi_4^y = \sig t,
&&
\eta_4 = 0 ;
\label{gb-symm4}
\\
&
\tau_5 = 0,
&&
\xi_5^x = 0,
&&
\xi_5^y = 0,
&&
\eta_5 = f_1(y+\sqrtsig t) + f_2(y-\sqrtsig t)  .
\label{gb-symm5}
\end{align}
Symmetries \eqref{gb-symm1}--\eqref{gb-symm3} are translations,
and symmetry \eqref{gb-symm4} is a boost in the $(y,t)$-plane. 
Symmetry \eqref{gb-symm4} is a linear combination of $2$ infinite-dimensional families,
corresponding to the general solution of $P_{tt}-\sig P_{yy}=0$ for $P(y,t)$.
(ii) 
Additional point symmetries are admitted by the 2D generalized Boussinesq potential equation \eqref{gb-pot}
only when $p=1$:
\begin{equation}\label{gb-symm6}
\tau_6 = 2t,
\quad
\xi_6^x = x,
\quad
\xi_6^y = 2y,
\quad
\eta_6 = -(v+x) . 
\end{equation}
This symmetry is a scaling combined with a shift in $v$.  
\end{thm}

The corresponding symmetry transformation groups for arbitrary $p\neq0$ are respectively given by 
\begin{align}
(x,y,t,v)\to & 
(x+\epsilon,y,t,v) , 
\\
(x,y,t,v)\to &
(x,y+\epsilon,t,v) , 
\\
(x,y,t,v)\to &
(x,y,t+\epsilon,v) , 
\\
(x,y,t,v)\to &
(x,\cosh(\epsilon\sqrtsig)y + \sinh(\epsilon\sqrtsig)t, \cosh(\epsilon\sqrtsig)t + \sinh(\epsilon\sqrtsig)t, v), 
\\
(x,y,t,v)\to &
(x,y,t,f_1(y+\sqrtsig t) + f_2(y-\sqrtsig t)+v) ,
\end{align}
where $\epsilon\in\Rnum$ is the group parameter. 
The symmetry transformation group generated by the scaling-shift \eqref{gb-symm6}
in the case $p=1$ is given by 
\begin{equation}\label{gb-scaling}
(x,y,t,v)\to 
(e^{\epsilon}x,e^{2\epsilon}y,e^{2\epsilon}t,e^{-\epsilon}v -\tfrac{1}{2}e^{\epsilon}x) .
\end{equation}

\subsection{Variational symmetries}

We next determine which of the symmetries in Theorems~\ref{pointsymms-gkp} and~\ref{pointsymms-gb}
are variational. 

Recall, 
an infinitesimal symmetry $\mathrm{\hat X}=P\partial_v$ is a variational symmetry 
if it leaves invariant any given Lagrangian $L$ up to a total divergence, 
\begin{equation}\label{varL}
\mathrm{\hat X}L =D_t \Psi^t +D_x \Psi^x + D_y \Psi^y
\end{equation}
where $\Psi^t,\Psi^x,\Psi^y$ are functions of $t$, $x$, $y$, $v$, and derivatives of $v$. 
Typically, 
this invariance condition is checked by first computing $\mathrm{\hat X}L$
and then attempting to use integration by parts to bring this expression into the form of a total divergence. 
A much more efficient method can be used, 
which is based on the Euler operator (i.e.\ the variational derivative)
\begin{equation}
\begin{aligned}
E_v & = \partial_v -D_t \partial_{v_t} -D_x \partial_{v_x} -D_y \partial_{v_y} 
+ D_t^2 \partial_{v_{tt}} + D_x^2 \partial_{v_{xx}} + D_y^2 \partial_{v_{yy}} 
\\&\qquad
+ D_tD_x \partial_{v_{tx}} + D_tD_y \partial_{v_{ty}} + D_xD_y\partial_{v_{xy}} 
+\cdots . 
\end{aligned}
\end{equation}
This operator has the property that it annihilates a function of $t$, $x$, $y$, $v$, and derivatives of $v$
iff the function is equal to a total divergence $D_t \Psi^t +D_x \Psi^x + D_y \Psi^y$. 
Consequently, 
the variational symmetry condition \eqref{varL} can be formulated as
\begin{equation}\label{EulervarL}
E_v(\mathrm{\hat X}L)=0 . 
\end{equation}
Moreover, this condition can be further simplified \cite{Olv,Anc2016} 
by use of the variational identity
\begin{equation}\label{ELident}
\mathrm{\hat X}L = E_v(L)\mathrm{\hat X}v + D_t \Phi^t +D_x \Phi^x + D_y \Phi^y
\end{equation}
which is derived by integration by parts,
where $E_v(L)$ is the variational derivative of $L$. 
Combining this identity \eqref{ELident} and the Euler-operator equation \eqref{EulervarL}, 
we see that the variational symmetry condition \eqref{varL} becomes
\begin{equation}\label{varsymm}
E_v(PE_v(L)) =0 . 
\end{equation}
This equation involves only the symmetry characteristic $P$ 
and the expression for the left-hand side of Euler-Lagrange equations $E_v(L)=0$ given by $L$ (which is unchanged if any total divergence is added to $L$). 

We will now apply the variational symmetry condition \eqref{varsymm}
to the point symmetries admitted by the gKP and 2D gB equations \eqref{gkp-pot} and \eqref{gb-pot},
where $L$ is the respective Lagrangian \eqref{gkp-L} and \eqref{gb-L}. 
A straightforward computation gives the following results. 

\begin{prop}\label{varsymms-gkp}
The variational point symmetries admitted by the generalized KP potential equation \eqref{gkp-pot}
are generated by 
the translation symmetries \eqref{gkp-symm1}--\eqref{gkp-symm3}, 
the rotation-boost symmetry \eqref{gkp-symm5}, 
and the infinite symmetry families \eqref{gkp-symm6} and \eqref{gkp-symm7}. 
The scaling symmetry \eqref{gkp-symm4} is variational when the nonlinearity power is $p=1$, 
which coincides with the case $f_3(t)=3t$, $f_4(t)=f_5(t)=0$ in the infinite family \eqref{gkp-symm7}.
\end{prop}

\begin{prop}\label{varsymms-gb}
The variational point symmetries admitted by the 2D generalized Boussinesq potential equation \eqref{gb-pot}
are generated by 
the translation symmetries \eqref{gb-symm1}--\eqref{gb-symm3}, 
the boost symmetry \eqref{gb-symm4}, 
and the infinite symmetry families \eqref{gb-symm5}. 
The scaling-shift symmetry \eqref{gb-symm6} is not variational for any nonlinearity power $p$. 
\end{prop}

\section{Conservation laws}\label{conslaws}

Conservation laws are of basic importance for nonlinear evolution equations 
because, for all solutions,  
they provide physical, conserved quantities as well as conserved norms

For the gKP and 2D gB equations \eqref{gkp-pot} and \eqref{gb-pot}, 
a local conservation law is a continuity equation
\begin{equation}\label{conslaw}
D_t T+D_x X+D_y Y=0
\end{equation}
holding for all solutions $v(x,y,t)$ of these respective equations, 
where $T$ is the conserved density, and $(X,Y)$ is the spatial flux,
which are functions of $t$, $x$, $y$, $v$, and derivatives of $v$. 
When solutions $v(x,y,t)$ are considered 
in a given spatial domain $\Omega\subseteq\Rnum^2$, 
every conservation law yields a corresponding conserved integral 
\begin{equation}\label{conservedquantity}
\mathcal{C}[v]= \int_{\Omega} T\,dx\,dy
\end{equation}
satisfying the global balance equation
\begin{equation}\label{globalconslaw}
\frac{d}{dt}\mathcal{C}[v]= -\int_{\partial\Omega} (X,Y)\cdot\hat\nvec\,ds
\end{equation}
where $\hat\nvec$ is the unit outward normal vector of the domain boundary curve $\partial\Omega$, 
and where $ds$ is the arclength on this curve. 
This global equation \eqref{globalconslaw} has the physical meaning that
the rate of change of the quantity \eqref{conservedquantity} on the spatial domain 
is balanced by the net outward flux through the boundary of the domain. 

A conservation law is locally trivial if, for all solutions $v(x,y,t)$ in $\Omega$,
the conserved density $T$ reduces to a spatial divergence $D_x \Psi^x + D_y \Psi^y$ 
and the spatial flux $(X,Y)$ reduces to a time derivative $-D_t(\Psi^x,\Psi^y)$, 
since then the global balance equation \eqref{globalconslaw} becomes an identity. 
Likewise, two conservation laws are locally equivalent 
if they differ by a locally trivial conservation law, for all solutions $v(x,y,t)$ in $\Omega$. 
We will be interested only in locally non-trivial conservation laws. 

Any non-trivial conservation law \eqref{conslaw} 
can be expressed in an equivalent characteristic form \cite{Olv,BluCheAnc,Anc2016}
which is given by a divergence identity holding off of the space of solutions $v(x,y,t)$. 
For the gKP and 2D gB equations, 
conservation laws have the respective characteristic forms
\begin{equation}\label{gkp-chareqn}
D_t\tilde T+D_x\tilde X+D_y\tilde Y=(v_{tx} +v_x^p v_{xx} +v_{xxxx} +\sig v_{yy})Q
\end{equation}
and
\begin{equation}\label{gb-chareqn}
D_t\tilde T+D_x\tilde X+D_y\tilde Y=(v_{tt}-v_{xx} -(v_x^{p+1})_x \mp v_{xxxx} -\sig v_{yy})Q
\end{equation}
where $Q$, $\tilde T$, $\tilde X$, $\tilde Y$ are functions of $t$, $x$, $y$, $v$, and derivatives of $v$,
and where the conserved density $\tilde T$ and the spatial flux $(\tilde X,\tilde Y)$
reduce to $T$ and $(X,Y)$ when restricted to all solutions $v(x,y,t)$ of the respective equations. 
These divergence identities are called the characteristic equation for the conservation law,
and the function $Q$ is called the conservation law multiplier. 
In particular, 
$Q$ has the property that it is non-singular when evaluated on any solution $v(x,y,t)$. 
Consequently, 
note that the characteristic equation of a conservation law is locally equivalent to the conservation law itself. 

Since the gKP equation and the 2D gB equation each possess a Lagrangian formulation, 
Noether's theorem \cite{Olv,BluCheAnc} shows that for both equations
a one-to-one correspondence holds between 
locally non-trivial conservation laws (up to equivalence) and variational symmetries. 
Specifically, this correspondence can stated simply as 
\begin{equation}
Q=P
\end{equation}
where $Q$ is a conservation law multiplier 
and $P$ is a variational symmetry characteristic. 
The conserved density $\tilde T$ and the spatial flux $(\tilde X,\tilde Y)$ 
corresponding to a given variational symmetry characteristic $P$
can be obtained straightforwardly by several methods \cite{Wol,BluCheAnc,Anc2016},
starting from the characteristic equations \eqref{gkp-chareqn} and \eqref{gb-chareqn}. 

For any given variational symmetry characteristic $P$, 
a repeated integration process \cite{Wol} 
can be directly applied to the terms in the expression $P E_v(L)$,
which yields $\tilde T,\tilde X,\tilde Y$. 
This method can sometimes be lengthy or awkward, 
depending on the complexity of this expression $P E_v(L)$,
A more systematic method \cite{BluCheAnc} 
consists of splitting the characteristic equation 
$D_t\tilde T+D_x\tilde X+D_y\tilde Y=P E_v(L)$ 
with respect to $v$ and its derivatives, 
giving an overdetermined linear system that can be solved to obtain $\tilde T,\tilde X,\tilde Y$. 
An alternative more direct method is to use a homotopy integral formula that inverts the Euler operator in the variational symmetry equation $E_v(P E_v(L))=0$. 
The simplest version of this formula appears in Refs.\cite{BluCheAnc,Anc2016};
a more complicated general version is given in Ref.\cite{Olv}. 

We will now summarize the conservation laws that arise from 
the variational point symmetries obtained in Propositions~\ref{varsymms-gkp} and~\ref{varsymms-gb}. 

\begin{thm}\label{conslaws-gkp}
(i)
The conservation laws corresponding to the variational point symmetries 
admitted by the generalized KP potential equation \eqref{gkp-pot} for arbitrary $p\neq 0$ 
are given by:
\begin{align}
& \begin{aligned}
T_1 = & \tfrac{1}{2}v_{xx}^2-\tfrac{1}{2}\sig v_y^2 -\tfrac{1}{(p+1)(p+2)}v_x^{p+2}
\\
X_1 = & v_t v_{xxx}-v_{tx}v_{xx}+\tfrac{1}{p+1} v_x^{p+1}v_t +\tfrac{1}{2}v_t^2
\\
Y_1= & \sig v_t v_y
\end{aligned}
\label{gkp-conslaw1}
\\\nonumber\\
& \begin{aligned}
T_2 = & \tfrac{1}{2}v_x^2
\\
X_2 = & v_xv_{xxx}-\tfrac{1}{2}v_{xx}^2+\tfrac{1}{p+2} v_x^{p+2} -\tfrac{1}{2}\sig v_y^2
\\
Y_2 = & \sig v_xv_y
\end{aligned}
\label{gkp-conslaw2}
\\\nonumber\\
& \begin{aligned}
T_3 = & \tfrac{1}{2}v_xv_y
\\
X_3 = & v_yv_{xxx}-v_{xx}v_{xy}+\tfrac{1}{p+1}v_y v_x^{p+1}+\tfrac{1}{2}v_t v_y
\\
Y_3 = & \tfrac{1}{2}v_{xx}^2+\tfrac{1}{2}\sig v_y^2 -\tfrac{1}{(p+1)(p+2)} v_x^{p+2} -\tfrac{1}{2}v_t v_x
\end{aligned}
\label{gkp-conslaw3}
\\\nonumber\\
& \begin{aligned}
T_4 = & \tfrac{1}{2}yv_x^2 -\sig tv_y v_x 
\\
X_4 = & (yv_x-2\sig t v_y)v_{xxx} -\tfrac{1}{2}yv_{xx}^2 + 2\sig t v_{xy}v_{xx} -\tfrac{1}{2}\sig y v_y^2  -\sig tv_yv_t 
\\&\qquad
-2\sig \tfrac{1}{p+1} tv_y v_x^{p+1} +\tfrac{1}{p+2}y v_x^{p+2}
\\
Y_4 = & -\sig tv_{xx}^2 -tv_y^2 +\sig yv_yv_x +\sig tv_tv_x +2\sig \tfrac{1}{(p+1)(p+2)} t v_x^{p+2} 
\end{aligned}
\label{gkp-conslaw4}
\\\nonumber\\
& \begin{aligned}
T_5 = & 0
\\
X_5 = & (\tfrac{1}{p+1}v_x^{p+1}+v_{xxx}+v_t )(f_1(t)y+ f_2(t))
\\
Y_5 = &  \sig ((yv_y-v) f_1(t)+ f_2(t)v_y)
\end{aligned}
\label{gkp-conslaw5}
\end{align}
(ii) 
The only additional conservation laws corresponding to variational point symmetries 
admitted by the generalized KP potential equation \eqref{gkp-pot} 
arise for $p=1$:
\begin{align}
& \begin{aligned}
T_6 = & \tfrac{1}{2}f_5(t) v_x^2 +f'_5(t)v
\\
X_6 = & \sig(f_5(t) v_x +\tfrac{1}{2}f''_5(t) y^2  -f'_5(t)x) v_{xxx}
-\tfrac{1}{2} f_5(t)v_{xx}^2 +f'_5(t)v_{xx}+\tfrac{1}{3}f_5(t) v_x^3 
\\&\quad
+(\tfrac{1}{4}\sig y^2 f''_5(t) -\tfrac{1}{2}f'_5(t)x )v_x^2
-\tfrac{1}{2}\sig f_5(t) v_y^2 +(\tfrac{1}{2}\sig f''_5(t) y^2 -f'_5(t)x) v_t 
\\
Y_6 = & \sig f_5(t) v_x v_y+(\tfrac{1}{2}f''_5(t) y^2 -\sig f'_5(t)x)v_y -f''_5(t)yv
\end{aligned}
\label{gkp-conslaw6}
\\\nonumber\\
& \begin{aligned}
T_7 = & -\tfrac{1}{4}\sig f'_4(t) y v_x^2  +\tfrac{1}{2} f_4(t)v_yv_x -\tfrac{1}{2}\sig f''_4(t) yv
\\
X_7 = & (-\tfrac{1}{2}\sig f'_4(t) yv_x + f_4(t)v_y -\tfrac{1}{12}f'''_4(t)y^3
+\tfrac{1}{2}\sig f''_4(t) yx )v_{xxx}
+\tfrac{1}{4}\sig f'_4(t) yv_{xx}^2  
\\&\quad
-( f_4(t)v_{xy} +\tfrac{1}{2}\sig f''_4(t)y)v_{xx}
-\tfrac{1}{6}\sig f'_4(t)yv_x^3 +(\tfrac{1}{2} f_4(t)v_y-\tfrac{1}{24}f'''_4(t)y^3 
\\&\quad
+\tfrac{1}{4}\sig f''_4(t) yx )v_x^2
+\tfrac{1}{4}f'_4(t)yv_y^2+\tfrac{1}{2} f_4(t)v_t v_y
\\&\quad
+(-\tfrac{1}{12}f'''_4(t)y^3+\tfrac{1}{2}\sig xy f''_4(t) )v_t 
\\
Y_7 = & \tfrac{1}{2} f_4(t)v_{xx}^2-\tfrac{1}{6} f_4(t)v_x^3 -(\tfrac{1}{2}f'_4(t)yv_y +\tfrac{1}{2} f_4(t)v_t )v_x 
+\tfrac{1}{2}\sig  f_4(t)v_y^2 
\\&\quad
+(-\tfrac{1}{12}\sig f'''_4(t)y^3 +\tfrac{1}{2}f''_4(t)yx)v_y
+(\tfrac{1}{4}\sig  f'''_4(t)y^2 -\tfrac{1}{2}f''_4(t) x)v
\end{aligned}
\label{gkp-conslaw7}
\\\nonumber\\
& \begin{aligned}
T_8 = & \tfrac{1}{2} f_3(t)v_{xx}^2 
-\tfrac{1}{6} f_3(t)v_x^3+\tfrac{1}{12}(2f'_3(t)x -\sig f''_3(t) y^2)v_x^2
+\tfrac{1}{3}f'_3(t)yv_yv_x 
\\&\quad
-\tfrac{1}{2}\sig f_3(t)v_y^2
+\tfrac{1}{6}(2f''_3(t)x -\sig f'''_3(t)y^2)v 
\\
X_8 = & ( (-\tfrac{1}{6}\sig f''_3(t) y^2 +\tfrac{1}{3}f'_3(t)x )v_x
+\tfrac{2}{3} f'_3(t) yv_y
+ f_3(t)v_t +\tfrac{1}{3}f'_3(t)v 
-\tfrac{1}{72}f''''_3(t)y^4 
\\&\quad
+\tfrac{1}{6}\sig f'''_3(t)y^2x -\tfrac{1}{6}f''_3(t)x^2 )v_{xxx} 
+( \tfrac{1}{12}\sig f''_3(t) y^2 -\tfrac{1}{6}f'_3(t)x )v_{xx}^2
\\&\quad
+( -\tfrac{2}{3}f'_3(t)y v_{xy}-\tfrac{2}{3}f'_3(t) v_x  -\tfrac{1}{6}\sig f'''_3(t)y^2
+\tfrac{1}{3}f''_3(t) x- f_3(t)v_{tx} )v_{xx}
\\&\quad
+( -\tfrac{1}{18}\sig f''_3(t) y^2 +\tfrac{1}{9}f'_3(t)x )v_x^3
+(\tfrac{1}{3}f'_3(t)yv_y+\tfrac{1}{2} f_3(t)v_t +\tfrac{1}{6}f'_3(t)v 
\\&\quad
-\tfrac{1}{144}f''''_3(t)y^4 +\tfrac{1}{12}\sig f'''_3(t) y^2x 
-\tfrac{1}{12}f''_3(t)x^2)v_x^2 
-\tfrac{1}{3}f''_3(t)v_x
\\&\quad
+(\tfrac{1}{12}f''_3(t) y^2 -\tfrac{1}{6}\sig f'_3(t)x )v_y^2
+\tfrac{1}{3}f'_3(t) yv_tv_y +\tfrac{1}{2}  f_3(t)v_t^2
\\&\quad
+(\tfrac{1}{3} f'_3(t)v -\tfrac{1}{72}f''''_3(t)y^4+\tfrac{1}{6}\sig f'''_3(t)y^2x 
-\tfrac{1}{6}f''_3(t)x^2)v_t 
\\
Y_8 = & \tfrac{1}{3}f'_3(t)y v_{xx}^2 
-\tfrac{1}{9} f'_3(t) yv_x^3 +( (-\tfrac{1}{6}f''_3(t) y^2 +\tfrac{1}{3}\sig f'_3(t)x )v_y
-\tfrac{1}{3}f'_3(t) yv_t )v_x 
\\&\quad
+\tfrac{1}{3}\sig f'_3(t)yv_y^2 
+(\sig  f_3(t)v_t +\tfrac{1}{3}\sig f'_3(t) v -\tfrac{1}{72}\sig f''''_3(t)y^4 +\tfrac{1}{6}f'''_3(t) y^2x 
\\&\quad
-\tfrac{1}{6}\sig f''_3(t)x^2 )v_y 
+(\tfrac{1}{18}\sig f''''_3(t)y^3-\tfrac{1}{3}f'''_3(t)yx )v 
\end{aligned}
\label{gkp-conslaw8}
\end{align}
\end{thm}

\begin{thm}\label{conslaws-gb}
(i)
The conservation laws corresponding to the variational point symmetries 
admitted by the 2D generalized Boussinesq potential equation \eqref{gb-pot}
for arbitrary $p\neq 0$ 
are given by:
\begin{align}
& \begin{aligned}
T_1 =&  \tfrac{1}{2}(v_t^2 \mp v_{xx}^2+v_x^2 +\sig v_y^2) + \tfrac{1}{p+2}v_x^{p+2} 
\\
X_1 =&  -(v_x^{p+1} \pm v_{xxx}+v_x)v_t \pm v_{tx}v_{xx}
\\
Y_1 =&  -\sig v_tv_y
\end{aligned}
\label{gb-conslaw1}
\\\nonumber\\
& \begin{aligned}
T_2 =&  v_xv_t
\\
X_2 =&  -\tfrac{p+1}{p+2} v_x^{p+2} \mp (v_xv_{xxx} -\tfrac{1}{2}v_{xx}^2) +\tfrac{1}{2}\sig v_y^2 -\tfrac{1}{2}v_t^2 -\tfrac{1}{2}v_x^2
\\
Y_2 =&  -\sig v_yv_x
\end{aligned}
\label{gb-conslaw2}
\\\nonumber\\
& \begin{aligned}
T_3 =&  v_yv_t
\\
X_3 =&  -(v_x^{p+1} \pm v_{xxx} +v_x)v_y \pm v_{xy}v_{xx}
\\
Y_3 =&  \tfrac{1}{p+2}v_x^{p+2} +\tfrac{1}{2} (\mp v_{xx}^2 +v_x^2 -\sig v_y^2-v_t^2)
\end{aligned}
\label{gb-conslaw3}
\\\nonumber\\
& \begin{aligned}
T_4 =&  \tfrac{1}{2} y(v_t^2 \mp v_{xx}^2 +v_x^2+\sig v_y^2) +\sig tv_yv_t + \tfrac{1}{p+2}yv_x^{p+2} 
\\
X_4 =&  -(t\sig v_y+yv_t)(\pm v_{xxx}+v_x+v_x^{p+1}) \pm (\sig tv_{xy}+yv_{tx})v_{xx}
\\
Y_4 =&  \tfrac{1}{p+2}tv_x^{p+2} -\sig(\tfrac{1}{2}(\pm v_{xx}^2+\sig v_y^2+v_t^2-v_x^2)t+yv_tv_y))
\end{aligned}
\label{gb-conslaw4}
\\\nonumber\\
& \begin{aligned}
T_5 =&  (f_1(y+\sqrtsig t)+f_2(y-\sqrtsig t))v_t -\sqrtsig (f'_1(y+\sqrtsig t)-f'_2(y-\sqrtsig t))v 
\\
X_5 =&  -( f_1(y+\sqrtsig t)+ f_2(y-\sqrtsig t))(\pm v_{xxx}+v_x+v_x^{p+1})
\\
Y_5=&  \sig( (f'_1(y+\sqrtsig t) + f'_2(y-\sqrtsig t))v 
-(f_1(y+\sqrtsig t) +f_2(y-\sqrtsig t))v_y )
\end{aligned}
\label{gb-conslaw5}
\end{align}
(ii)
No additional conservation laws corresponding to variational point symmetries 
admitted by the 2D generalized Boussinesq potential equation \eqref{gb-pot} 
arise for any $p\neq 0$. 
\end{thm}

\subsection{Conserved quantities of the gKP equation}

For the gKP potential equation, 
conservation law \eqref{gkp-conslaw1} arises from the time-translation symmetry \eqref{gkp-symm1} 
and yields the energy quantity
\begin{equation}\label{gkp-ener}
\mathcal{E}[v] = \int_{\Omega} \big( 
\tfrac{1}{2}v_{xx}^2-\tfrac{1}{2}\sig v_y^2 -\tfrac{1}{(p+1)(p+2)}v_x^{p+2}
\big) dx\,dy .
\end{equation}
Conservation laws \eqref{gkp-conslaw2} and \eqref{gkp-conslaw3}
arise from the space-translation symmetries \eqref{gkp-symm2} and \eqref{gkp-symm3},
both of which yield momentum quantities
\begin{align}
\mathcal{P}^x[v] & = \int_{\Omega} \tfrac{1}{2}v_x^2\, dx\,dy,
\label{gkp-x-mom}
\\
\mathcal{P}^y[v] & = \int_{\Omega} \tfrac{1}{2}v_xv_y\, dx\,dy .
\label{gkp-y-mom}
\end{align}
Conservation law \eqref{gkp-conslaw4} arises from the rotation-boost symmetry \eqref{gkp-symm4}
and yields an analogous momentum quantity
\begin{equation}\label{gkp-rotboostmom}
\mathcal{Q}[v] = \int_{\Omega} \big( \tfrac{1}{2}yv_x^2 -\sig tv_y v_x \big) dx\,dy .
\end{equation}
In addition to these conserved quantities, 
there are two spatial flux quantities which arise from conservation law \eqref{gkp-conslaw5} and describe conserved topological charges, 
\begin{align}
\mathcal{F}_1[v] & = \int_{\partial\Omega} \big( \tfrac{1}{p+1}v_x^{p+1}+v_{xxx}+v_t , \sig v_y \big)\cdot\hat\nvec\, ds
=0, 
\label{gkp-charg1}
\\
\mathcal{F}_2[v] & = \int_{\partial\Omega} \big( y(\tfrac{1}{p+1}v_x^{p+1}+v_{xxx}+v_t ), \sig (yv_y-v) \big)\cdot\hat\nvec\, ds
=0 . 
\label{gkp-charg2}
\end{align}

All of the preceding conserved quantities \eqref{gkp-ener}--\eqref{gkp-charg2}
hold for $p\neq0$. 
They are homogeneous under the scaling symmetry \eqref{gkp-scaling}. 
In particular, they have the respective scaling weights
\begin{equation}
w_{\mathcal{E}} = 1-4/p,
\quad
w_{\mathcal{P}^x} = 3-4/p, 
\quad
w_{\mathcal{P}^y} = 2-4/p, 
\quad
w_{\mathcal{Q}} = 5-4/p, 
\end{equation}
and 
\begin{equation}
w_{\mathcal{F}_1} = -2/p, 
\quad
w_{\mathcal{F}_2} = 2-2/p, 
\end{equation}
as defined by $\mathcal{C}[v]\to e^{w\epsilon} \mathcal{C}[v]$. 

Finally, 
the three infinite families of conservation laws \eqref{gkp-conslaw6}--\eqref{gkp-conslaw8}
which hold only when $p=1$ 
yield two conserved dilational momentum quantities 
\begin{align}
\mathcal{\tilde P}^x[v] & = \int_{\Omega} \big(  \tfrac{1}{2}f_5(t) v_x^2 +f'_5(t)v \big) dx\,dy ,
\label{gkp-dil-x-mom}
\\
\mathcal{\tilde P}^y[v] & = \int_{\Omega} \big( \tfrac{1}{2} f_4(t)v_yv_x -\tfrac{1}{4}\sig f'_4(t) y v_x^2  -\tfrac{1}{2}\sig f''_4(t) yv \big) dx\,dy ,
\label{gkp-dil-y-mom}
\end{align}
and a conserved dilational energy quantity
\begin{equation}
\begin{aligned}
\mathcal{\tilde E}[v] & = \int_{\Omega} \big( 
\tfrac{1}{2} f_3(t)( v_{xx}^2 -\sig v_y^2 -\tfrac{1}{3} v_x^3 )
+\tfrac{1}{3} f'_3(t)( \tfrac{1}{2} x v_x^2 +yv_yv_x )
\\&\qquad
+\tfrac{1}{3} f''_3(t)( x v -\tfrac{1}{4} y^2v_x^2 )
-\tfrac{1}{6}\sig f'''_3(t)y^2 v 
\big) dx\,dy ,
\end{aligned}
\label{gkp-dil-ener}
\end{equation}
all of which are not scaling homogeneous
unless the functions $f_3,f_4,f_5$ are chosen to be monomials in $t$. 

The conserved energies \eqref{gkp-ener} and \eqref{gkp-dil-ener}, 
conserved momenta \eqref{gkp-y-mom}--\eqref{gkp-rotboostmom}, \eqref{gkp-dil-x-mom}--\eqref{gkp-dil-y-mom}, 
and topological charges \eqref{gkp-charg1}--\eqref{gkp-charg2}
each exhibit an essential dependence on the potential $v$. 
Consequently, the corresponding conservation laws are nonlocal in terms of $u$. 

The only conservation law that is local in terms of $u$ is the $x$-momentum, 
whose corresponding conserved quantity \eqref{gkp-x-mom} describes 
the mass of $u$:
\begin{equation}
\mathcal{M}[u]= \int_{\Omega} u^2\, dx\,dy = 2\mathcal{P}^x[v] .
\end{equation}
This quantity is scaling invariant iff $p=\tfrac{4}{3}$. 
In comparison, the energy is scaling invariant iff $p=4$.

\subsection{Conserved quantities of the 2D gB equation}

Similarly, for the gB potential equation, 
conservation law \eqref{gb-conslaw1} yields the energy quantity
\begin{equation}\label{gb-ener}
\mathcal{E}[v] = \int_{\Omega} \big( 
\tfrac{1}{2}(v_t^2 \mp v_{xx}^2+v_x^2 +\sig v_y^2) + \tfrac{1}{p+2}v_x^{p+2} 
\big) dx\,dy 
\end{equation}
which arises from the time-translation symmetry \eqref{gb-symm1}. 
Conservation laws \eqref{gb-conslaw2} and \eqref{gb-conslaw3}
yield the momentum quantities
\begin{align}
\mathcal{P}^x[v] & = \int_{\Omega} v_xv_t\, dx\,dy,
\label{gb-x-mom}
\\
\mathcal{P}^y[v] & = \int_{\Omega} v_yv_t \, dx\,dy ,
\label{gb-y-mom}
\end{align}
which arise from the space-translation symmetries \eqref{gb-symm2} and \eqref{gb-symm3}. 
Conservation law \eqref{gb-conslaw4} yields the boost-momentum quantity 
\begin{equation}\label{gb-boostmom}
\mathcal{Q}[v] = \int_{\Omega} \big( 
\tfrac{1}{2} y(v_t^2 \mp v_{xx}^2 +v_x^2+\sig v_y^2) +\sig tv_yv_t + \tfrac{1}{p+2}yv_x^{p+2}
\big) dx\,dy,
\end{equation}
arising from the boost symmetry \eqref{gb-symm4}. 
The infinite family of conservation laws \eqref{gb-conslaw5} 
yield a corresponding infinite family of quantities
\begin{equation}\label{gb-transverse-mom}
\mathcal{T}[v] = \int_{\Omega} \big( 
(f_1(y+\sqrtsig t)+f_2(y-\sqrtsig t))v_t -\sqrtsig (f'_1(y+\sqrtsig t)-f'_2(y-\sqrtsig t))v 
\big) dx\,dy 
\end{equation}
which are counterparts of transverse momentum quantities 
known for the linear wave equation $v_{tt}-\sig v_{yy} =0$. 
In particular, 
if we put $f_1=f_2=\tfrac{1}{2}$
then the conserved quantity $\mathcal{T}[v] = \int_{\Omega} v_t\, dx\,dy = \dot A[v]$
is the time derivative of the net amplitude $A[v]= \int_{\Omega} v\, dx\,dy$.
This implies $\ddot A[v]=0$ and hence 
$A[v]= t\dot A[v]|_{t=0} + A[v]|_{t=0}$. 
Thus, the net amplitude has the same motion as a free particle. 
If we next put 
$f_1=\tfrac{1}{2}(y+\sqrtsig t)$ and $f_2=\tfrac{1}{2}(y-\sqrtsig t)$,
then we obtain the conserved quantity
$\mathcal{T}[v] = \int_{\Omega} yv_t\, dx\,dy = \dot A_1[v]$
which is the time derivative of the first $y$-moment of the net amplitude, 
$A_1[v] = \int_{\Omega} yv\, dx\,dy$. 
This implies $\ddot A_1[v]=0$ and hence 
$A_1[v]= t\dot A_1[v]|_{t=0} + A_1[v]|_{t=0}$. 
If we then put 
$f_1=\tfrac{1}{2}(y+\sqrtsig t)^2$ and $f_2=\tfrac{1}{2}(y-\sqrtsig t)^2$,
the resulting conserved quantity is 
$\mathcal{T}[v] = \int_{\Omega} ( (y^2+\sig t^2)v_t -2\sig t v) dx\,dy = \dot A_2[v] +\sig t^2\dot A[v] -2\sig t A[v]$,
where $A_2[v] = \int_{\Omega} y^2v\, dx\,dy$ is the second $y$-moment of the net amplitude.
Hence we get $\ddot A_2[v]= 2\sig A[v]$,
which yields $A_2[v]= \tfrac{1}{3}\sig t^3 \dot A[v]|_{t=0} + \sig t^2 A[v]|_{t=0} + t\dot A_2[v]|_{t=0} + A_2[v]|_{t=0}$. 
Similar expressions arise for the higher $y$-moments $A_n[v] = \int_{\Omega} y^n v\, dx\,dy$, 
$n=1,2,\ldots$. 

The preceding conserved quantities \eqref{gb-ener}--\eqref{gb-transverse-mom}
hold for $p\neq0$. 
They have no scaling properties unless $p=1$,
since this is the only nonlinearity power for which a scaling symmetry \eqref{gb-symm6} exists. 
This scaling symmetry contains a shift on $v$ and generates the transformation group \eqref{gb-scaling}. 
The derivatives of $v$ can be shown to transform in a scaling homogeneous manner, 
except for $v_x\to e^{-2\epsilon} v_x -\tfrac{1}{2}$ which contains a shift. 

To work out how the conserved quantities will transform,
we can apply a general result \cite{AncKar} that 
relates the symmetry action on conservation laws 
to the commutators of the scaling symmetry and the variational symmetries 
which generate the conservation laws. 
We find that the energy \eqref{gb-ener}, $y$-momentum \eqref{gb-y-mom}, and boost momentum \eqref{gb-boostmom}
are each scaling homogeneous, 
modulo the addition of a locally trivial conserved quantity. 
In particular, they have the respective scaling weights 
\begin{equation}
w_{\mathcal{E}} = w_{\mathcal{P}^x} =w_{\mathcal{P}^y} = -3,
\quad
w_{\mathcal{Q}} = -1 ,
\end{equation}
as defined by $\mathcal{C}[v]\to e^{w\epsilon} \mathcal{C}[v]$. 
In contrast, we find that the $x$-momentum \eqref{gb-x-mom} 
is only scaling homogeneous modulo the addition of 
the conserved transverse momentum quantity $\dot A[v]$
as well as a locally trivial conserved quantity. 
Moreover, 
the infinite family of quantities \eqref{gb-transverse-mom} is not scaling homogeneous 
unless the functions $f_1,f_2$ are chosen to be monomials in $y\pm \sqrtsig t$. 

Finally, 
all of these conserved quantities \eqref{gb-ener}--\eqref{gb-transverse-mom}
exhibit an essential dependence on the potential $v$,
and so the corresponding conservation laws are nonlocal in terms of $u$.

\section{Line soliton solutions}\label{solns}

A line soliton is a solitary travelling wave 
\begin{equation}\label{linesoliton}
u=U(x+\mu y-\nu t) 
\end{equation}
in two dimensions,
where the parameters $\mu$ and $\nu$ determine the direction and the speed of the wave. 
A more geometrical form for a line soliton is given by writing 
$x+\mu y= (x,y)\cdot\kvec$ 
with $\kvec=(1,\mu)$ being a constant vector in the $(x,y)$-plane. 
The travelling wave variable can then be expressed as 
\begin{equation}
\xi = x+\mu y-\nu t = |\kvec|( \hat\kvec\cdot (x,y) - c t )
\end{equation}
where the unit vector 
\begin{equation}
\hat\kvec = (\cos\theta,\sin\theta),
\quad
\tan\theta = \mu
\end{equation}
gives the direction of propagation of the line soliton,
and the constant 
\begin{equation}
c = \nu/|\kvec|, 
\quad
|\kvec|^2 = 1+\mu^2
\end{equation}
gives the speed of the line soliton. 
Here the direction obeys $-\pi/2\leq\theta\leq \pi/2$,
while the speed can be $c>0$ or $c<0$. 
Note that a line soliton \eqref{linesoliton} is invariant 
under the two-parameter group of translations
$t\to t + \epsilon_1$, $x\to x + \epsilon_1 \nu -\epsilon_2 \mu$, $y\to y + \epsilon_2$,
with $\epsilon_1,\epsilon_2\in\Rnum$. 

We will now derive the explicit line soliton solutions \eqref{linesoliton}
for the gKP equation \eqref{gKP} and the 2D gB equation \eqref{gB}. 
It will be convenient to use the coordinate form of the travelling wave variable 
$\xi = x+\mu y-\nu t$ for this derivation. 

Substitution of the line soliton expression \eqref{linesoliton}
into both the gKP and 2D gB equations in potential form 
yields a nonlinear third-order ODE. 
To reduce these respective ODEs to separable ODEs which can be integrated, 
we need two functionally-independent first integrals. 
The necessary first integrals can be obtained by the following steps 
from the conservation laws given in Theorems~\ref{conslaws-gkp} and~\ref{conslaws-gb}. 

Suppose a conservation law $D_t T+D_x X+D_y Y=0$ 
does not contain the variables $t,x,y$ explicitly. 
Then the conservation law gives rise to a first integral of the line soliton ODE 
by the reductions 
\begin{equation}\label{reduced-derivatives}
D_t\big|_{u=U(\xi)} = -\nu \frac{d}{d\xi},
\quad
D_x\big|_{u=U(\xi)} = \frac{d}{d\xi},
\quad
D_y\big|_{u=U(\xi)} = \mu\frac{d}{d\xi},
\end{equation}
yielding
\begin{equation}\label{reduced-conslaw}
\frac{d}{d\xi} \Big( (X +\mu Y - \nu T)\big|_{u=U(\xi)} \Big) = 0. 
\end{equation}
The first integral is thus given by 
\begin{equation}\label{firstintegral}
X +\mu Y - \nu T = C = \const 
\end{equation}
It has the physical meaning of the spatial flux of the conserved quantity $\int_{\Omega} T\, dx\,dy$ 
in a reference frame moving with speed $c$ in the direction $\kvec$ in the $(x,y)$ plane
(namely, the rest frame of the line soliton).

\subsection{gKP line solitons}

We begin with the gKP potential equation \eqref{gkp-pot}. 
Using the relation \eqref{pot} for $u$ in terms of the potential $v$, 
and substituting the line soliton expression \eqref{linesoliton}, 
we obtain the nonlinear third-order ODE 
\begin{equation}\label{gkp-U-ode}
(\sig \mu^2 -\nu)U' +U^p U' +U''' = 0
\end{equation}
for $U(\xi)$. 
From Theorem~\ref{conslaws-gkp}, 
we see that the gKP conservation laws that do not explicitly contain $t,x,y$ consist of 
the energy \eqref{gkp-conslaw1}, 
the $x$ and $y$ momenta \eqref{gkp-conslaw2}--\eqref{gkp-conslaw3}, 
and the spatial flux \eqref{gkp-conslaw5} with $f_2=1$ and $f_1=0$. 
When the first integral formula \eqref{firstintegral} is applied to these four conservation laws, 
we find that, up to a constant proportionality factor, 
the energy and the $y$-momentum yield the same first integral as the $x$-momentum, 
\begin{equation}\label{gkp-FI1}
C_1 = UU'' +\tfrac{1}{2}(\sig\mu^2-\nu)  U^2 -\tfrac{1}{2} U'^2+\tfrac{1}{p+2} U^{p+2} ,
\end{equation}
while the spatial flux yields an independent first integral, 
\begin{equation}\label{gkp-FI2}
C_2 = U'' +(\sig\mu^2-\nu)U +\tfrac{1}{p+1} U^{p+1} .
\end{equation}
We are interested in a solitary wave solution, 
and so we impose the asymptotic conditions
$U,U',U''\to 0$ as $|\xi|\to \infty$. 
This requires $C_1=C_2=0$. 
Then, combining the two first integrals \eqref{gkp-FI1} and \eqref{gkp-FI2}, 
we obtain the first-order separable ODE
\begin{equation}
U'{}^2 = (\nu -\mu^2\sig)U^2 -\tfrac{2}{(p+2)(p+1)} U^{p+2}
\end{equation}
which can be straightforwardly integrated. 
Its general solution, up to a shift in $\xi$, is given by 
\begin{equation}\label{gkp-linesoliton}
U(\xi) = (2(p+2)(p+1)\kappa)^{1/p} \sech\big(\tfrac{1}{2}p(p+1)(p+2)\sqrt{\kappa}\,\xi\big)^{2/p},
\quad
\kappa = \nu -\mu^2\sig
\end{equation}
which is smooth and satisfies the desired decay conditions if $\kappa >0$ and $p>0$. 
This is the general line soliton solution of the gKP equation \eqref{gKP} for all $p>0$. 
It coincides with the well-known line soliton solution \cite{Sat} for the ordinary KP equation \eqref{KP} when $p=1$. 

We next discuss a few properties of this solution \eqref{gkp-linesoliton}. 

First, the parameters $\mu=\tan\theta$ and $\nu=c\sqrt{1+\mu^2}$ 
give the angle $\theta$ of the direction of motion of the line soliton with respect to the $x$ axis, 
and the speed $c$ of the line soliton along its direction of motion. 
These two parameters need to obey the kinematic condition 
\begin{equation}
\nu > \sgn(\sig) \mu^2 
\end{equation}
where, recall, $\sig=\pm 1$. 
Since $c$ must be finite, we conclude that the line soliton can propagate 
in any direction except $|\theta|=\pi/2$ (namely, parallel to the $y$ axis).
In the case $\sig = -1$, 
the kinematic condition is satisfied for all positive speeds $c>0$ 
as well as for negative speeds $0>c>-\mu^2/\sqrt{1+\mu^2}$. 
In contrast, in the case $\sig= 1$, 
the kinematic condition requires that the speed is larger than $c>\mu^2/\sqrt{1+\mu^2}>0$ and hence must be positive. 
These kinematical properties are independent of the nonlinearity power. 

Last, 
the physical conserved quantities for the line soliton are given by 
the $x$ and $y$ momenta densities
\begin{equation}
(\tfrac{1}{2}q\kappa)^{1+2/p}\sech\big(\tfrac{1}{2}pq\sqrt{\kappa}\xi\big)^{4/p},
\quad
\mu (\tfrac{1}{2}q\kappa)^{1+2/p}\sech\big(\tfrac{1}{2}pq\sqrt{\kappa}\xi\big)^{4/p}
\end{equation}
and the energy density 
\begin{equation}
(\tfrac{1}{2}q\kappa)^{1+2/p}\sech\big(\tfrac{1}{2}pq\sqrt{\kappa}\xi\big)^{4/p} ((q+1/q)\tanh\big(\tfrac{1}{2}pq\sqrt{\kappa}\xi\big)^2 -q\nu/\kappa)
\end{equation}
where $q=(p+1)(p+2)$. 
These densities are obtained from the conservation law expressions 
\eqref{gkp-conslaw1}--\eqref{gkp-conslaw3}, 
evaluated for the line soliton solution \eqref{gkp-linesoliton}.

\subsection{2D gB line solitons}

We apply the preceding derivation to the 2D gB potential equation \eqref{gb-pot}. 

The nonlinear third-order ODE for the line soliton \eqref{linesoliton} is given by 
\begin{equation}\label{gb-U-ode}
(1+\sig \mu^2 -\nu^2)U' +(p+1)U^p U' \pm U''' = 0 . 
\end{equation}
From Theorem~\ref{conslaws-gb}, 
we see that the 2D gB conservation laws that do not explicitly contain $t,x,y$ consist of 
the energy \eqref{gb-conslaw1}, 
the $x$ and $y$ momenta \eqref{gb-conslaw2}--\eqref{gb-conslaw3}, 
and the transverse momentum \eqref{gb-conslaw5} with $f_1=f_2=1$. 
The first integral formula \eqref{firstintegral} then yields 
\begin{equation}\label{gb-FI1}
C_1= \pm U'' +(1+\mu^2\sig -\nu^2)U +U^{p+1}
\end{equation}
which arises from the transverse momentum, 
and also 
\begin{equation}\label{gb-FI2}
C_2= \pm(U U'' -\tfrac{1}{2}U'^2) +\tfrac{1}{2}(1+\mu^2\sig -\nu^2)U^2 +\tfrac{p+1}{p+2} U^{p+2} 
\end{equation}
which comes from the energy as well as from both the $x$ and $y$ momenta.
These two first integrals are functionally independent. 
Since we are interested in a solitary wave solution, 
and we impose the asymptotic conditions
$U,U',U''\to 0$ as $|\xi|\to \infty$,
which require $C_1=C_2=0$. 
Then the two first integrals \eqref{gb-FI1} and \eqref{gb-FI2} can be combined to obtain 
the first-order separable ODE
\begin{equation}
U'{}^2 = \pm( (\nu^2 -\mu^2\sig-1)U^2 -\tfrac{2}{p+2} U^{p+2} )
\end{equation}
which is straightforward to integrate. 
Its general solution, up to a shift in $\xi$, is given by 
\begin{equation}\label{gb-linesoliton}
U(\xi) = (\tfrac{1}{2}(p+2)\kappa)^{1/p} \sech\big(\tfrac{1}{2}p(p+2)\sqrt{\pm\kappa}\,\xi\big)^{2/p},
\quad
\kappa = \nu^2 -\mu^2\sig -1 . 
\end{equation}
This solution will be smooth and satisfy the desired decay conditions 
if $\pm\kappa >0$ and if either $p>0$ in the ``$+$'' case 
or if $1/p$ is a rational number with an odd denominator in the ``$-$'' case. 

Hence, when $\kappa >0$, 
this solution is a line soliton of the ``$+$'' case of the 2D gB equation for all $p>0$, 
whereas when $\kappa <0$, 
it is a line soliton of the ``$-$'' case of the 2D gB equation 
provided $p$ is a positive rational number with an odd numerator. 

We next discuss a few properties of this solution \eqref{gb-linesoliton}. 
Recall that $\mu$ and $\nu$ give 
the angle $\theta=\arctan\mu$ of the direction of motion of the line soliton with respect to the $x$ axis, 
and the speed $c=\nu/\sqrt{1+\mu^2}$ of the line soliton. 
Since $c$ must be finite, note that the direction of motion cannot be parallel to the $y$ axis. 

First, 
in the ``$+$'' case of the 2D gB equation, 
the parameters need to obey the kinematic condition 
\begin{equation}
\nu^2 > \sgn(\sig) \mu^2 +1 
\end{equation}
where, recall, $\sig=\pm 1$. 
This kinematic condition requires that the speed is $|c|>1$ 
in the case $\sig= 1$,
and $|c|>\sqrt{(1-\mu^2)/(1+\mu^2)}$ 
in the case $\sig=-1$. 
Hence when the direction of motion is within an angle of $\pi/4$ with respect to the $x$-axis in the case $\sig =-1$, 
there is a minimum speed $|c|> \sqrt{(1-\mu^2)/(1+\mu^2)} >0$. 

Second, 
in the ``$-$'' case of the 2D gB equation, 
the line soliton parameters $\mu$ and $\nu$
need to obey the opposite kinematic condition 
\begin{equation}
\nu^2 < \sgn(\sig) \mu^2 +1 . 
\end{equation}
Consequently, in the case $\sig=1$, 
there is a maximum speed $|c|< 1$. 
In the case $\sig=-1$, 
the direction of motion is confined to within an angle of $\pi/4$ with respect to the $x$-axis,
with the speed obeying $|c|<\sqrt{(1-\mu^2)/(1+\mu^2)}$. 
These kinematical properties are independent of the nonlinearity power. 

Last, 
the physical conserved quantities for the line soliton are given by 
the $x$ and $y$ momenta densities
\begin{equation}
-\nu(\tfrac{1}{2}q\kappa)^{1/p}\sech\big(\tfrac{1}{2}pq\sqrt{\pm\kappa}\xi\big)^{4/p},
\quad
-\mu\nu (\tfrac{1}{2}q\kappa)^{1+2/p}\sech\big(\tfrac{1}{2}pq\sqrt{\kappa}\xi\big)^{4/p}
\end{equation}
and the energy density 
\begin{equation}
-(\tfrac{1}{2}q\kappa)^{1+2/p}\sech\big(\tfrac{1}{2}pq\sqrt{\pm\kappa}\xi\big)^{4/p} ((q+1/q)\tanh\big(\tfrac{1}{2}pq\sqrt{\pm\kappa}\xi\big)^2 -2q\nu^2/\kappa)
\end{equation}
where $q=p+2$. 
These densities are obtained from the conservation law expressions 
\eqref{gb-conslaw1}--\eqref{gb-conslaw3}, 
evaluated for the line soliton solution \eqref{gb-linesoliton}.

\section{Concluding remarks}\label{remarks}

For the generalized KP equation \eqref{gKP} and the 2D generalized Boussinesq equation \eqref{gB}
with $p$-power nonlinearities, 
we have derived all conservation laws that arise from variational point symmetries
and used them to obtain the general line soliton solutions of these two equations 
for all powers $p>0$. 
We have also discussed some basic kinematical properties of the line solitons. 

Our results can be further used as a starting point to investigate the stability of the line soliton solutions for the gKP equation. 
It is known that \cite{KadPet,OKeiPar} 
the line solitons of the KP equation ($p=1$) are 
stable in the case $\sig =1$ and unstable in the opposite case $\sig =-1$. 
An interesting question is to determine the stability of the line solitons for higher powers $p\geq 2$. 
Compared to the KdV equation, 
in two dimensions the critical power for the conserved mass is $p=\tfrac{4}{3}$,
while the critical power for the conserved energy is $p=4$. 

In addition to line soliton solutions, 
the KP equation possesses lump solutions, 
which are rational functions and involve several parameters. 
No lump solutions are known for the modified KP equation, 
and an interesting question is to see if a similar result will hold 
for the gKP equation for all $p>1$. 

Stability of line solitons and existence of lump solutions can be investigated similarly 
for the 2D gB equation. 

In a different direction, 
there has been recent work \cite{MatWatTan}
investigating solutions of a combined KP and modified KP equation
$(u_t+6\alpha u u_x+6 \beta u^2u_x+u_{xxx})_x+u_{yy}=0$
and also a combined 2D Boussinesq and modified 2D Boussinesq equation 
$u_{tt}=u_{xx}+(6\alpha u u_x+6 \beta u^2u_x+u_{xxx})_x+u_{yy}$. 
A natural problem will be to look for 
line solitons, lump solutions, and other types of solitary waves, 
as well as conservation laws, 
of the fully generalized KP and 2D Boussinesq equations
$(u_t+f(u) u_x+u_{xxx})_x+\sig u_{yy}=0$
and 
$u_{tt}=u_{xx}+(f(u) u_x+u_{xxx})_x+\sig u_{yy}$.

\section*{Acknowledgements}
S.C.A.\ is supported by an NSERC research grant. 
Takayuki Tschudia is thanked for helpful remarks.

\end{document}